\documentstyle[epsfig]{l-aa}   

\def \rsun {\ifmmode$R$_{\odot}\else R$_{\odot}$\fi}

\def \hcm {\hbox {\ifmmode $ H atoms cm$^{-2}\else H atoms cm$^{-2}$\fi}}
\def\approxgt{\mathrel{\hbox{\rlap{\lower.55ex \hbox {$\sim$}}
        \kern-.3em \raise.4ex \hbox{$>$}}}}
\def\approxlt{\mathrel{\hbox{\rlap{\lower.55ex \hbox {$\sim$}}
        \kern-.3em \raise.4ex \hbox{$<$}}}}

\newcommand {\sax} {{\it BeppoSAX }}

\def\lsim{\lower.5ex\hbox{$\; \buildrel < \over \sim \;$}}
\def\gsim{\lower.5ex\hbox{$\; \buildrel > \over \sim \;$}}

\begin{document}

   \title{BeppoSAX discovery of a new Seyfert 2 galaxy: 1SAXJ2234.8-2541
\thanks{Partially based on observations collected at the European Southern 
Observatory, Chile }}


   \author{A. Malizia\inst{1,2}, L. Bassani\inst{3}, I. Negueruela\inst{1}}

   \offprints{A. Malizia}

\institute{
   {\inst{1}\sax Science Data Center, ASI,
    Via Corcolle, 19,
    I-00131 Roma , Italy}\\
    {\inst{2}Department of Physics and Astronomy,
     Southampton  University
     Southampton, SO17 1BJ, England}\\
   {\inst{3}Istituto TESRE, CNR,
    Via Gobetti, 101,
    I-40129 Bologna, Italy}\\
}
   \date{Received  ; accepted }

   \maketitle
   \markboth{Malizia et al., }{}

\begin{abstract}
In the present work we report the \sax serendipitous discovery of the
type 2 AGN 1SAXJ2234.8-2541 in the MECS field of view when pointing at
the Seyfert 2 galaxy NGC7314. The source is optical identified with
the bright (m$_{B}$=14.40) galaxy ESO533-G50 at redshift $z$=0.026.
The source is clearly detected at energies above 4 keV but barely
visible below this energy implying heavy obscuration intrinsic to the
source.  Spectral analysis indicates a column density of the order of
2-3 $\times$ 10$^{23}$ cm$^{-2}$ and a power law photon index
compatible with values often seen in active galaxies.  These X-ray
characteristics suggest a Seyfert 2 classification for
1SAXJ2234.8-2541.  Subsequent spectroscopy of the optical conterpart,
ESO533-G50, performed with the ESO 1.52m telescope at La Silla
Observatory, confirms the type 2 nature of this source and therefore
its identification with the X-ray source.  We also report the marginal
detection of an iron line centered at 7.29 $\pm$ 0.47 keV and having
an equivalent width of 464$^{\rm+636}_{\rm-398}$ eV; although
marginal, this result is indicative of the presence of warm
material in the source.  
If instead the line is associated to cold material, we estimate an upper
limit to its equivalent width of 290 eV.
The overall characteristics of 1SAXJ2234.8-2541 strongly suggest that 
the source is Compton thin.

\end{abstract}

\begin{figure*}
\psfig{file=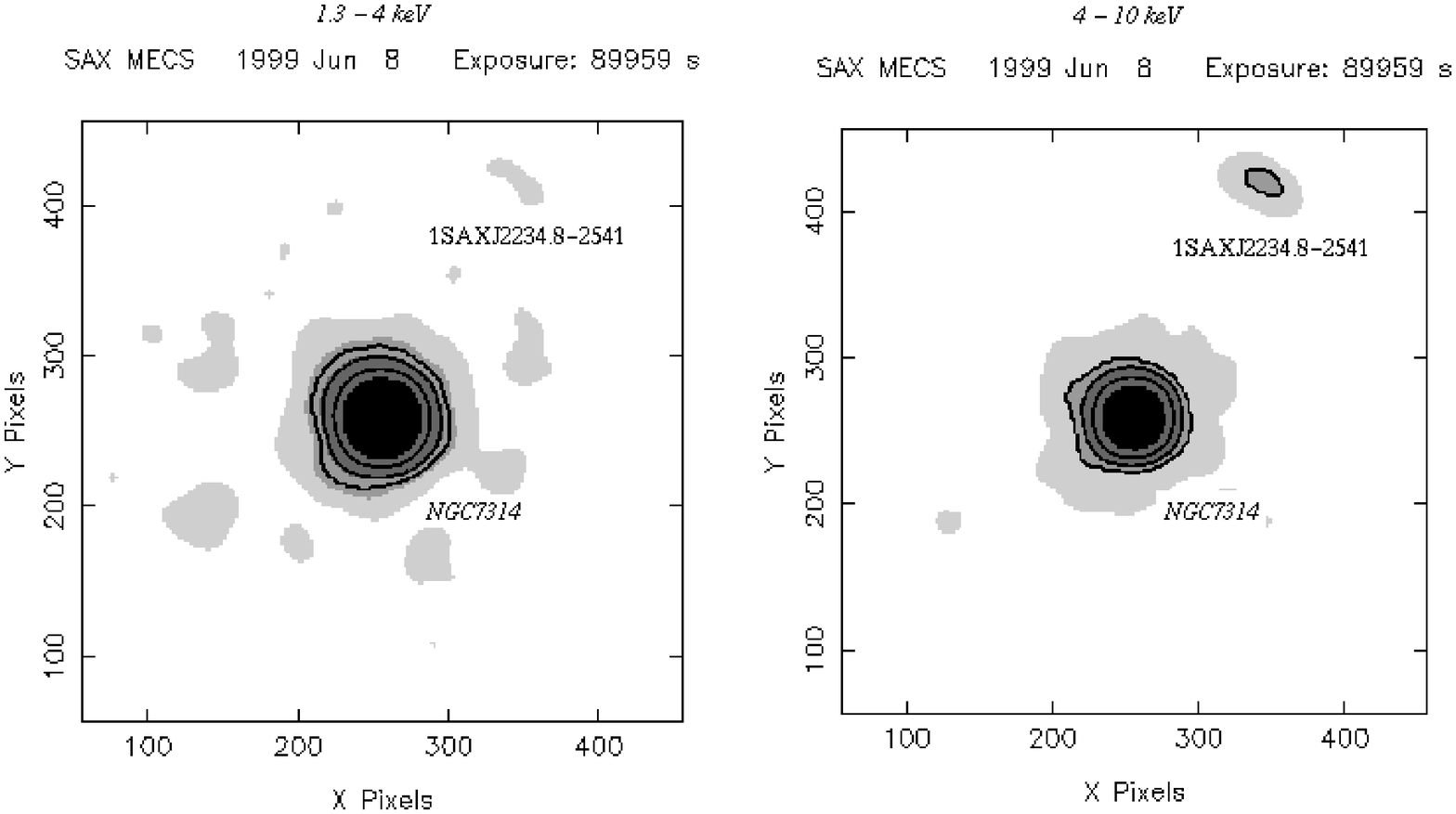,width=16cm,height=8cm}
\caption{\sax MECS image when pointing at NGC7314 in two
different energy bands: 1.3-4 keV (left side) and 4-10 keV 
(right side).}
\end{figure*}

\section{Introduction}
It is rather clear that in the soft X-ray band (0.5-2 keV), the major
contribution to the Cosmic X-Ray Background (XRB) is due to the
superposition of many discrete sources which are basically AGN
(Hasinger et al. 1998, Schmidt et al. 1998).  On the contrary, due to
the lack, until recent years, of sensitive imaging instrument at
higher energies (above 2 keV), the nature of the objects responsible
of the hard (2-50 keV) XRB remains still dubious.  Furthermore at
higher energy the XRB shows a thermal-like spectrum (KT $\sim$ 40 keV)
difficult to explain with any class of sources known so far.  However,
the current suggestion is that even at higher energies the main
contributors to XRB are the AGN, but this time the heavy obscured
ones.  This is in line with both the theoretical expectations (Setti
and Woltjier, 1989; Matt and Fabian, 1994; Comastri et al. 1995) and
the observational findings (Bassani et al. 1999; Fiore et al. 1999).
A great contribution to this issue has been given both by ASCA (Della
Ceca et al. 1999, Boyle et al. 1998, Ueda et al. 1998) and \sax
(Giommi et al. 1998a,1998b, Fiore et al. 1999) and more recently 
by Chandra (Mushotzky et al 2000).  In
particular the good sensitivity of the \sax MECS detectors (5-10 keV
flux limit of $\sim$ 0.002 mCrab in 100Ks, Boella et al. 1997a) and
their improved point spread function (PSF), provide an ideal tool for
discovering hard X-ray sources, i.e. those with emission above 4 keV.
A step ahead in this context has been made with the \sax High Energy
LLarge Area Survey (HELLAS) (Fiore et al. 1998a,1998b).  
The HELLAS survey has so far catalogued 180 sources (Fiore et al. 1999)
in about 50
deg$^{2}$ of sky down to a limiting flux of 5 $\times$ 10$^{-14}$ erg
cm$^{-2}$ s$^{-1}$.  Most of the sources are identified
with AGN which corresponds to resolving about 30\% of the hard XRB.
Furtermore, a great majority of them turned out to be absorbed by
column densities in the range 10$^{22}$-10$^{23}$ cm$^{-2}$ even
though they may be highly different in their optical and near-infrared
properties.  Fabian (2000) has recently pointed out that if absorbed
AGN are the major contributors to the hard XRB then about 85$\%$ of
the accreation power in the Universe is absorbed.  Identification of
hard X-ray sources and study of their X-ray characteristics is
therefore important in order to investigate all the above mentioned
issues.  In this paper we report the \sax serendipitous discovery of a
type 2 AGN: 1SAXJ2234.8-2541. Its absorbed type 2 nature was firstly
revealed by the X-ray data and after confirmed by the optical
spectroscopy.  We anticipate that the new generation of X-ray
observatories like Chandra and XMM will discover many more sources
similar to 1SAXJ2234.8-2541 and few others recently reported in the
literature (Della Ceca et al. 1999, Fiore et al. 1999) making these
results reference cases for future work.

\section{X-Ray Data Analysis}

The \sax X-ray observatory (Boella et al. 1997b) is a major
programme of the Italian Space Agency with participation of the
Netherlands Agency for Aereospace Programs.  This work concerns
results obtained with the Medium Energy Concentrator Spectrometers
(MECS; Boella et al. 1997a). The data from the LECS
(Low Energy Concentrator Spectrometer) were
not available whereas the data from PDS (Phoswich Detector System) 
cannot be used due to contamination
from the main target of the observation which was pointed by the
BeppoSAX Narrow Field Instruments (NFI) 
from June 8th to June 10th, 1999.

\subsection{Imaging Analysis}
The reduction procedures and screening criteria used to produce th
linearized and equalized (between the two MECS) event files were
standard (Guainazzi et al. 1999) and took into account the offset
position of the source.\\ 1SAXJ2234.8-2541 was discovered in the MECS
field of view when pointing at the Seyfert 2 galaxy NGC7314.  Figures
1 shows the MECS image in two different bands (1.3-4 keV) and (4-10
keV) with overlaid a contour plot indicating the intensity level of
the X-ray emission.  The hard X-ray nature of 1SAXJ2234.8-2541 is
clearly evident: the source is almost undetectable in the soft X-ray
band on the left side of the figure while it is well detected at 
a confidence level of 10 $\sigma$ in the hard X-ray band 
on the right side.  There are no
pointed ROSAT observation containing this source, nor is this object
detected in the ROSAT all sky survey.  In the X-ray band the source is
located at (J2000) 
RA = 22$^{h}$ 34$^{m}$ 53.6$^{s}$ Dec = -25{$^{\circ}$} 
41$^{'}$ 2.6$^{''}$, 25 arcmin
north-east of the target source; the uncertainty associated with the
source position is 1 arcmin (90\% confidence level).  A fairly bright
(m$_{B}$ = 14.40) spiral SB galaxy is present in this error circle and
is identified with ESO533-G50 at redshift $z$=0.026.  Figure 2 shows
the ESO Digitized Sky Survey image centered on ESO533-G50 with
superimposed  the X-ray positition of 1SAXJ2234.8-2541 
and the associated error circle.  
No much data are available on this source from the
literature, although the galaxy is catalogued as ringed (Buta et
al. 1995, see also figure 2) probably as a result of a galaxy-galaxy
and/or intergalactic gas-galaxy collisions.  The probability of finding
such a bright galaxy within an error box of 1 arcmin is $\sim$ 1
$\times$ 10$^{-3}$ (Giommi et al. 2000).  This combined with the data
discussed in this paper strongly support the identification of
1SAXJ2234.8-2541 with the ESO galaxy.

\begin{figure}
\psfig{file=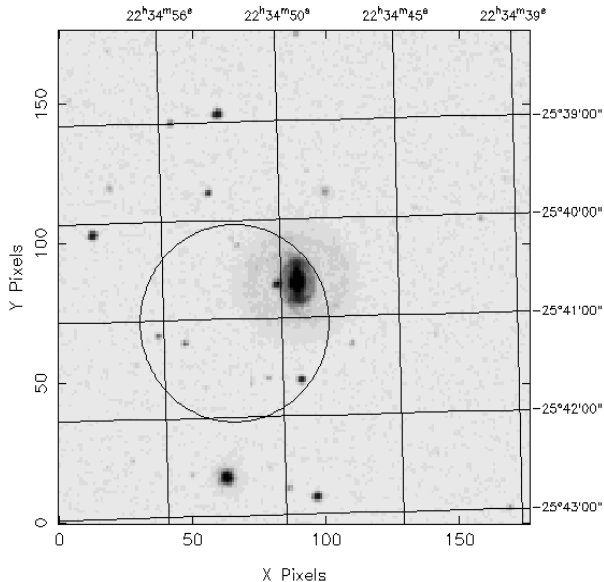,width=8cm,height=8cm}
\caption{DSS image of the field centered on ESO533-050.
The circle represents the 90$\%$ confidence level error box of 1 arcmin
radius while the center of the circle is the X-ray position of 
1SAXJ2234.8-2541.}
\end{figure}

\subsection{Spectral Analysis}

The effective on-source exposure time was 89959 s.  Spectral data were
extracted from regions centered on 1SAXJ2234.8-2541 with a radius of
4$'$ and the background subtraction was performed by means of blank
sky spectra extracted from the same region of the source.  The net
source count rates were (3.11$\pm$0.26)$\times10^{-3}$ cts/s in the
(1.8 - 10.5 keV) MECS energy range.

MECS data were rebinned in order to sample the energy resolution of
the detector with an accuracy proportional to the count rate.  The
spectral analysis has been performed by means of the {\sc XSPEC 10.0}
package, and using the instrument response matrices released by the
BeppoSAX Science Data Centre in September 1997.  All quoted errors
correspond to 90\% confidence intervals for one interesting parameter
($\Delta\chi^2$ = 2.71).  Source plus background light curves did not
show any significant variation and therefore the data from the whole
observation were summed together for the spectral analysis.

\begin{figure}
\psfig{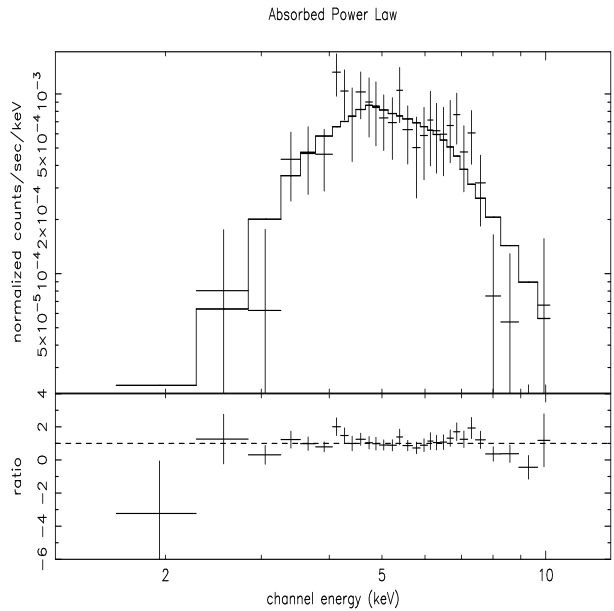}
\caption{BeppoSAX MECS data of 1SAXJ2234.8-2541 fitted with
absorbed power law (top); the residuals between the data and the model
are plotted at the bottom.}
\end{figure}

\begin{figure}
\psfig{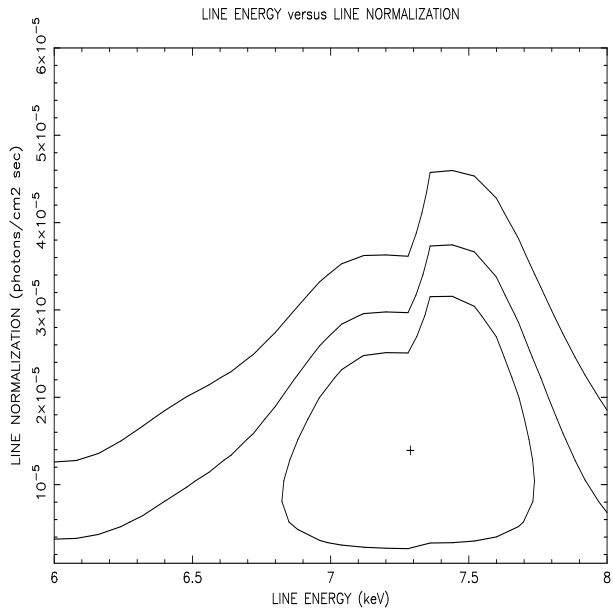}
\caption{BeppoSAX confidence contours of the normalization of
the line vs the Gaussian line. The contours are 68\%, 90\% and
98\% confidence level.}
\end{figure}

All the models used in what follows contain an additional term to
allow for the absorption of X-rays due to our Galaxy that in the
direction of 1SAXJ2234.8-2541 amounts to $1.5\times10^{20}$ cm$^{-2}$
(Dickey \& Lockman, 1990).  We first fit the data
with a single power law to search for any extra feature: this fit
gives an unusually flat spectrum ($\Gamma$ = +0.7) and marginally
acceptable reduced $\chi^{2}$ = 2.3 for 28 d.o.f..
We also try a thermal fit
using a Raymond Smith model but also this gives a too high temperature
(greater than 10 keV) and an unacceptable fit.  We then introduced
intrinsic absorption in the source: in this case the fit is
satisfactory ($\chi^{2}/\nu$=21.32/27) and results in a spectrum
having a photon index $\Gamma$=2.98$^{+1.58}_{-1.21}$ and an absorbing
column density N$_{\rm H}$=(3.1$^{\rm+1.5}_{\rm-1.2}$) $\times$
10$^{23}$ cm$^{-2}$.  The introduction of this extra parameter provides
an improvement in the fit ($\Delta\chi^{2}$=43.9 for one additional
parameter) which is significant at more than 99.99$\%$ confidence
level.  If the power law index is fixed to 1.9, a value more
appropriate to an AGN X-ray spectrum, the column density reduces to
2.2 $\times$ 10$^{23}$ cm$^{-2}$ and the $\chi^{2}$ is 23.41 for 28
d.o.f.  The 2-10 keV observed flux is 1.5 $\times$ 10$^{-12}$ erg
cm$^{-2}$ s$^{-1}$ corresponding to an absorption corrected luminosity
of 1.2 $\times$10$^{43}$ erg s$^{-1}$. The source count rate spectrum
and the data to model ratio for the absorbed power law fit with index
fixed to 1.9, are shown in figure 3.  Inspection of figure 3 indicates
that some residual emission may be present around 6-8 keV suggesting
the introduction in the fit of a narrow gaussian emission line.  The
line turns out to be centered at 7.29$\pm$0.47 keV and has a rest frame
equivalent width (EW) of 464$^{\rm+636}_{\rm-398}$ eV.  The addition
of this line provides a slight improvement in the fit
($\Delta\chi^{2}$=3.5 for two additional parameters) which is only
significant at 87.5$\%$ confidence level. 
If the line width is allowed to vary, the best fit value is 
0.28$^{+0.54}_{-0.28}$ i.e. consistent with being zero.  
The confidence contours of the line energy versus normalization 
(figure 4) indicate that the line is also compatible 
with a K$_{\alpha}$ line from cold material at 6.4 keV.
However, fixing the line energy at 6.4 keV
provides an upper limit of 290 eV to the equivalent width of 
this line and does not eliminate the residuals observed in 
the data to model ratio in figure 3.
We therefore conclude that the observed line, if confirmed,
is more likely associated to warm than cold material.

\section{Optical Data Analysis}
In order to optically classify our source by applying the standard
line ratio diagnostic (see for example Tresse et al. 1996),
spectroscopy covering 3400-5400 $\AA$ region is needed.  Therefore we
performed such a type of observation as soon as was possible.  Optical
data were obtained on the night of November 2, 1999, using the Boller
\& Chivens spectrograph on the ESO 1.52-m telescope at the La Silla
Observatory (Chile). The spectrograph was equipped with the
holographic grating \#33 and the Loral \#39 CCD camera. The
configuration results in a nominal dispersion of $\sim
1$\AA/pixel. Data have been reduced according to the standard
procedure using the {\em Starlink} software packages {\sc ccdpack}
(Draper 1998) and {\sc figaro} (Shortridge et al. 1997). The extracted
spectrum is displayed in figure 5.  Unfortunately conditions were not photometric
and therefore no flux calibration was attempted.  \\ 
The spectrum is typical of a low-redshift galaxy, except for the strong emission lines
corrisponding to forbidden transitions of oxygen. H$\beta$ cannot be
seen in emission at the expected wavelength of $\lambda4491\AA$
implying that it could be extremely weak or absorbed.
The fact that the only strong lines that we can detect are forbidden,
strongly suggests that ESO533-G050 is a type 1.9-2 Seyfert
(see Osterbrock 1981 and Osterbrock and Dahari 1983 for similar 
examples).
Because we cannot see H$\beta$, we cannot formally apply any of 
the diagnostic ratios, such as its position in the 
O[{\sc iii}]$\lambda5007$/H$\beta$
against O[{\sc ii}]$\lambda$3727/H$\beta$ diagram.  However, in such a
diagram, Seyfert 2 galaxies occupy the top-right part, with high
values of both O[{\sc iii}]/H$\beta$ and O[{\sc ii}]/H$\beta$.
1SAXJ2234.8-2541 seems to have extremely high values for these two
indices and therefore should be classified along with these sources
in agreement with the X-ray spectroscopic results.

\begin{figure}
\psfig{file=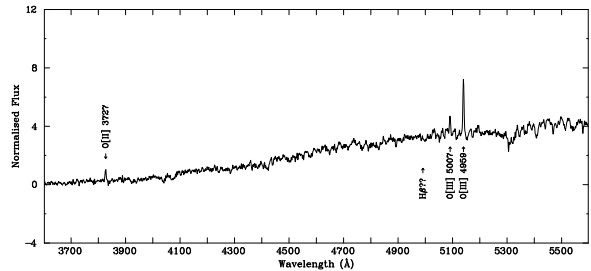,width=8cm, bbllx=80,bblly=350,bburx=500,bbury=535}
\caption{Optical spectrum of 1SAXJ2234.8-2541 in the 3400-5400 $\AA$}
\end{figure}

\section{Discussion}

Having assessed that our object is a type 2 Seyfert, the next step is 
to determine what kind of Seyfert 2 it is, i.e. Compton thin or Compton
thick. The column density measured exclude that the source is thick
(i.e. the intrinsic N$_{H}$ is less than 
1.5 $\times$ 10$^{24}$ cm$^{-2}$). 
However, observations of Seyfert 2 limited to a restricted energy 
band such as is the 2-10 keV band, has  provided wrong estimates 
of the column density in a few occasions in the past 
(Cappi et al. 1999a, Turner et al 2000, Vignali et al. 2000). 
Therefore an independent
method to determine the nature of the hidden nucleus must be used:
following Mulchaey et al. (1994) the IR to X-ray ratio  is a valid
indicator of the Compton thinness/thickness of the source.
Although  the source was not detected by IRAS, we estimated
an upper limit to the (25-60 micron) flux using data obtained within
1 degree from the source and applying Mulchaey et al.
(1994) formulation. The value obtained (2.3 $\times$ 10$^{-11}$
erg cm$^{-2}$ s$^{-1}$)  provides a log(F$_{IR}$/F$_{X}$) 
of $\le$ 1.7 and, since in Compton thin cases this
value is expected to be $\sim$0.9, our result support the  Compton
thinness of the source.
Compton thin Seyfert 2 galaxies are generally characterized 
by an iron line emission produced either by transmission and/or 
reflection by cold absorbing material in the source, i.e. by a 
dominant line at 6.4 keV.
In the first case, we expect an equivalent width of 100-200 eV for a
column density of 3 $\times$ 10$^{23}$ while in the case of reflection
the width increases to about 400 eV (Turner et al. 1997) for 
the same amount of absorption. 
Our upper limit of 290 eV on the 6.4 K$\alpha$ line
marginally discriminate between these two possibilities suggesting 
that the reflection in this source is negligible or absent; 
also our spectrum is too steep
to allow for a strong reflection component which tends to flatten the
photon  index to a value smaller than 1.9.
On the other hand, the line could be at $\sim$7 keV, thus likely to 
come from a highly ionized medium in the source.
Ionized lines are generally observed in Starburst galaxies and Liners
(Cappi et al. 1999b
and Pellegrini et al. 1999, 2000 , Terashima et al. 1999) and
also in Compton thick Seyfert 2, but are rarely seen in Compton thin
type 2 objects (only 20-30$\%$ of the sources in the ASCA sample
analyzed by Turner et al. 1997 show iron lines from ionized
material and of these the majority are indeed Compton thick sources).
A line at 7.06 keV could be due to Fe K$\beta$ fluorescence 
associated to the 6.4 keV line; however the expected 
K$\beta$/K$\alpha$ ratio is 1:9 while in our
data the K$\beta$ line would dominate over the K$\alpha$ one. 
Obviously a deeper observation of 1SAXJ2234.8-2541
is necessary to confirm the presence of the line and its energy; 
this may be worth given the paucity of Compton thin Seyfert 2 
having warm material at the nucleus and the serendipitous 
discovery of the source, which makes it relevant to the
synthesis of the hard XRB.

\begin{acknowledgements}
This research has made use of SAXDAS linearized
and cleaned event files produced at the BeppoSAX Science Data Centre.
We thank F. Fiore and P. Giommi for useful discussion and suggestions.
IN is supported by an ESA external research fellowship.
\end{acknowledgements}

\end{document}